\documentclass[letterpaper, 10 pt, conference]{ieeeconf}  % ECC template (PaperCept ieeeconf.cls)
\IEEEoverridecommandlockouts  % needed for \thanks
\usepackage{amsmath,amssymb}
\usepackage{graphicx}
\usepackage{booktabs}
\usepackage{multirow}
\usepackage{cite}
\usepackage{xcolor}
\usepackage[hidelinks]{hyperref}
\graphicspath{{figs/}}
\newcommand{\Pel}{P_\mathrm{el}}
\newcommand{\Pref}{P_\mathrm{ref}}
\newcommand{\mc}{\dot m_\mathrm{c}}
\title{\LARGE \bf
Power Management of Electrically Driven Centrifugal Compressors for Flexible Operation
}

\author{Alp Albay$^{*}$, Tak Lam Fok$^{*}$, Niklaus Hung$^{*}$, Maiken Omtveit$^{\dagger,\ddagger}$, Mehmet Mercang\"oz$^{*}$%
\thanks{$^{*}$ Autonomous Industrial Systems Laboratory, Department of Chemical Engineering, Imperial College London, London, United Kingdom. }
\thanks{$^{\dagger}$ R\&D Automation and Electrification, ABB AS, Fornebu, Norway. }
\thanks{$^{\ddagger}$Department of Electric Energy, Norwegian University of Science and Technology, Trondheim, Norway. }
\thanks{This work has been submitted to the IEEE for possible publication. Copyright may be transferred without notice, after which this version may no longer be accessible.
}% TODO: e-mail ({\tt\small ...}), co-authors, funding acknowledgement
}

\begin{document}

\maketitle
\thispagestyle{empty}
\pagestyle{empty}

\begin{abstract}
We present a control strategy for power management of a variable-speed 15\,MW (rated power) centrifugal compressor, and investigate how quickly it can follow a reduced electrical power reference without entering surge. We use a Greitzer--Gravdahl type model scaled to industrial size with rate-limited valve actuators as a simulation environment and propose an advanced regulatory control structure using feedback control loops with dynamically scaled feedforward mechanisms and a structure of min/max selectors. In simulated experiments, a 6\,MW power reduction from varying initial operating points on a 12\,MW iso-power line is delivered at rates of up to 10\,MW/s, i.e. to 90\,\% in 0.54\,s, without opening the recycle valve. Depending on the initial operating point, a large power reduction can be delivered at once or with a restricted ramp rate. The stroking rate of the discharge side valve is identified as the limiting factor for how fast the power can be reduced in the studied configuration. A limited sensitivity study on the impact of rotor inertia for power shedding performance is also provided.
\end{abstract}
\section{Introduction}

Electrification of gas compression involves replacing gas-turbine drivers with variable-speed electric motors, both onshore and offshore~\cite{bh_icl}. A large electrically driven compressor of 10--20 \, MW, is a substantial and controllable load. It can support grid operation for example through participation in frequency services or, in islanded systems such as offshore platforms, reduce its demand when a generating unit trips. Industrial demand-side management has been studied extensively from a scheduling perspective~\cite{zhang2016}, but less in terms of fast response services. In these cases, it must be known how quick the power consumption of a compressor can be safely reduced while the machine keeps delivering gas.

When turning down the power consumption of a compressor quickly, the limiting phenomenon is surge. A power reduction decelerates the rotor, the head falls with the square of the speed, and unless the discharge pressure falls with it the operating point moves towards the surge line. Surge dynamics are well described by the Greitzer model~\cite{greitzer1976} and its extensions~\cite{moore1986,gravdahl1999}, and drive-torque actuation has been used for active surge control~\cite{gravdahl2002}. Industrial practice relies on a surge control line with a margin and a recycle, or anti-surge, valve~\cite{botros1994}. Model predictive control alternatives that coordinate speed and recycle have been demonstrated experimentally~\cite{cortinovis2015}. For large electric-driven compressors, the same coordination of drive and recycle has been used to ride through grid voltage dips, during which the drive torque drops suddenly~\cite{besselmann2017,cortinovis2019}. On islanded plants, compressor drives are typical candidates in contingency-based fast load shedding, which disconnects whole loads within 100--250 \, ms of a generator trip~\cite{petriccioli2021}. Tripping a compressor this way loses production, often causes flaring, and can require a full restart of the system. Thus, a partial reduction through a variable speed drive is a preferred alternative if it can be made fast and safe. However, care must be taken as opening the recycle valve is also undesirable as the drive then spends power recirculating gas, defeating the purpose of the reduction, and a wide opening interrupts the delivery of gas to the downstream process.

In this study, we use a dynamic model of a 15 MW electrically driven compressor to determine how fast its electrical power can be reduced on request without entering surge and, preferably, without recycling gas. We propose a regulatory control structure for this purpose, which coordinates the drive torque, the discharge valve and the anti-surge valve through feedforward elements and min/max selectors, and we test it with large power reductions from the operating points closest to surge, for both grid-service and island load-shedding requests. The contributions are: (i) a regulatory structure whose power mode combines a torque feedforward with a discharge-valve feedforward scaled with the size of the desired power reduction and a minimum-flow limit with a rate-scaled dynamic margin, protected by speed overrides in external-reset form and by anti-surge control with a recycle trip line, with the request clipped to the reachable minimum power (Section~\ref{sec:control}); (ii) an assessment of how fast a large power reduction can be delivered from the start points closest to surge, which shows that a rate limit of up to 10\,MW/s keeps the recycle valve closed while a step does not, and identifies the stroking rate of the discharge valve as the binding constraint (Section~\ref{sec:results}); (iii) the sensitivity of this result to drive lag, signaling delay and rotor inertia, assessed against grid and island requirements (Section~\ref{sec:results}).

\section{Process Model and Operating Envelope}\label{sec:model}

\subsection{Model}

\begin{figure}[t]
\centering
\includegraphics[width=\columnwidth]{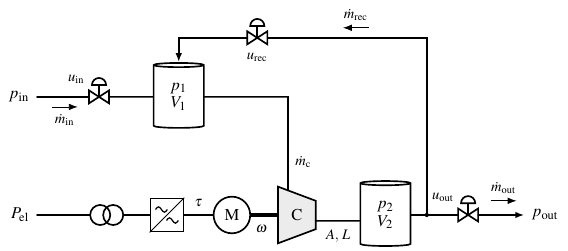}
\caption{Compression system: suction valve and plenum, motor-driven centrifugal compressor, discharge plenum and discharge valve to the header, and the recycle line with the anti-surge valve.}
\label{fig:process}
\end{figure}

The compression system (Fig.~\ref{fig:process}) has a suction plenum, a centrifugal stage driven by a variable-frequency drive and motor, a discharge plenum, a discharge (control) valve to a header, and a recycle line with an anti-surge valve. The states are the plenum pressures $p_1, p_2$, the compressor mass flow $\mc$, a lagged pressure-ratio state $\mathrm{PR}$ and the shaft speed $\omega$, governed by Equations \eqref{eq:p1}--\eqref{eq:rotor}. Equations \eqref{eq:p1} and \eqref{eq:p2} are mass balances of the suction and discharge plenums, where $a$ is the speed of sound and $V_1, V_2$ are the plenum volumes. Equation \eqref{eq:duct} is the momentum balance of the gas in the compressor duct of length $L$ and area $A$. The compressor flow accelerates when the pressure it produces, $\mathrm{PR}\,p_1$, exceeds the discharge pressure $p_2$, and decelerates, or reverses in surge, otherwise. 
\begin{align}
\dot p_1 &= \frac{a^2}{V_1}\left(\dot m_\mathrm{in} + \dot m_\mathrm{rec} - \mc\right), \label{eq:p1}\\
\dot p_2 &= \frac{a^2}{V_2}\left(\mc - \dot m_\mathrm{rec} - \dot m_\mathrm{out}\right), \label{eq:p2}\\
\frac{\mathrm{d}\mc}{\mathrm{d}t} &= \frac{A}{L}\left(\mathrm{PR}\,p_1 - p_2\right), \label{eq:duct}\\
\dot{\mathrm{PR}} &= \frac{1}{\tau_c}\left(\Pi(\omega,\mc,p_1) - \mathrm{PR}\right), \label{eq:pr}\\
J\dot\omega &= \tau_\mathrm{d} - \tau_\mathrm{load}. \label{eq:rotor}
\end{align}
In \eqref{eq:pr} the pressure ratio follows the steady-state map $\Pi$ with a time constant $\tau_c$, as in the Greitzer model~\cite{greitzer1976}. The map $\Pi$ is written in dimensionless form, head coefficient $\psi = h_s/U^2$, against flow coefficient $\varphi=\mc/(\rho_1 U D^2)$, as a Moore--Greitzer cubic~\cite{moore1986} whose coefficients depend on the tip Mach number. The cubic polynomials were fitted to a measured map, with an error of 1\,\% in pressure ratio. Surge is at the peak of each speed line. $h_s$ is the isentropic head, the specific enthalpy rise of an isentropic compression from $p_1$ to $\mathrm{PR}\cdot p_1$, $D$ is the impeller tip diameter, $U = \omega D/2$ the tip speed and $\rho_1$ the suction density. In \eqref{eq:rotor}, $J$ is the rotor inertia, $\tau_\mathrm{d}$ is the drive torque that accelerates the rotor, and $\tau_\mathrm{load}$ is the load torque that decelerates it. The load torque is the power absorbed by the gas divided by the speed, $\tau_\mathrm{load} = P_\mathrm{gas}/\omega$, plus windage and bearing losses of 1.5\,\% of the rated power at rated speed. The gas power is $P_\mathrm{gas} = \mc h_s/\eta$, where $\eta$ is the isentropic efficiency, which depends on the operating point and peaks at 0.88. Valve flows follow $\dot m = C u \sqrt{\rho\,|\Delta p|}$ with openings $u_\mathrm{in}, u_\mathrm{out}, u_\mathrm{rec}$; the recycle flow is taken from the discharge line after the cooler and is algebraic in its opening.

The electrical power drawn from the grid, which includes the losses of the motor, the variable-frequency drive, and the transformer, lumped into an efficiency $\eta_\mathrm{md}$, is:
\begin{equation}
\Pel = (\tau_\mathrm{d}\,\omega)/\eta_\mathrm{md}, \qquad \eta_\mathrm{md}=0.955, \label{eq:pel}
\end{equation}
so $\Pel$ exceeds the thermodynamic compression power by 6.5--7.4\,\% across the operating range. The drive works in torque control with a limit of limited to 1.1 times the rated torque. The torque produced by the motor follows the command instantaneously, except in Section~\ref{sec:lag}, where a first-order lag with time constant $\tau_\mathrm{e}$ is added. Methane is treated as a real gas, with $Z$ and $\kappa$ calculated at suction conditions using CoolProp~\cite{coolprop} and assumed constant. The model and the measured map originate from an experimental compressor test rig~\cite{cortinovis2015}. The model is scaled by similarity laws to a 15\,MW machine at 12\,000\,rpm (Table~\ref{tab:machine}), preserving the dimensionless map, the tip speed and the Greitzer $B$ parameter. The rotor inertia is set through the inertia constant $H$, the kinetic energy at rated speed divided by the rated power; published motor--compressor trains suggest 0.4--1.7\,s, and $H=1$\,s is used as nominal value.

Of the three valves in the model, the suction valve is not used for control and is fully open unless stated otherwise. The discharge and anti-surge valve actuators are modelled as pure rate limits. The discharge valve is a process control valve and strokes at 30\,\%/s (3.3\,s for full travel). The anti-surge valve is a fast-stroking valve and opens fully in 2\,s; such valves are specified for full opening in 1--2\,s~\cite{emerson_asv}. The model is integrated with RK4 at 1.7\,ms; the controllers run at 10\,ms, with measurement noise and first-order filters (pressures 2.5\,kPa, flows 0.5\,kg/s and power 15\,kW, $1\sigma$).

\begin{table}[t]
\caption{The 15\,MW machine}\label{tab:machine}
\centering\small
\begin{tabular}{lr}
\toprule
Rated power & 15\,MW\\
Speed range & 9.9--12\,krpm\\
Suction / header & 15.1 / 22.4\,bar\\
Impeller diameter & 0.57\,m\\
Tip speed & 360\,m/s\\
Rated torque & 11.94\,kN\,m\\
Inertia constant $H$ & 1\,s\\
Plenum volumes & 188 / 18.8\,m$^3$\\
Discharge valve rate limit & 30\,\%/s\\
Anti-surge valve rate limit & 50\,\%/s open, 2\,\%/s close\\
\bottomrule
\end{tabular}
\end{table}

\subsection{Operating envelope}
\begin{figure}[t]
\centering
\includegraphics[width=\columnwidth]{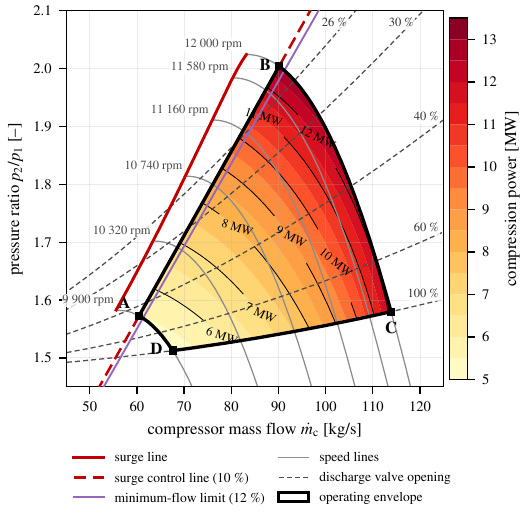}
\caption{Compressor map and operating envelope (A--B: surge control line, 10\,\%; B--C: maximum speed; C--D: discharge valve fully open; D--A: minimum speed), with the minimum-flow limit used in power mode (12\,\%). Colour: thermodynamic compression power.}
\label{fig:envelope}
\end{figure}

The operating envelope of the compressor is shown in Fig.~\ref{fig:envelope}. With suction open and recycle closed, the envelope is bounded by 4 lines: the surge control line (SCL) A--B, maximum speed B--C, the discharge valve fully open C--D, and minimum speed D--A. The SCL is a straight line through the 10\,\% flow-margin points at minimum and maximum speed; its margin to the true surge line is 9.4--10.5\,\%. At a given pressure ratio, the compressor surges if its flow falls to the surge flow $\dot m_\mathrm{s}(\mathrm{PR})$, the flow on the surge line at that pressure ratio. The surge margin
\begin{equation}
\mu = \frac{\mc}{\dot m_\mathrm{s}(\mathrm{PR})} - 1
\label{eq:margin}
\end{equation}
is the fraction by which the compressor flow is above the surge flow, $\dot m_\mathrm{s}$. Thus, at $\mu = 0$, the flow is on the surge line, and $\mu = 0.1$ means 10\,\% more flow than at surge, which is approximately the SCL. Sustained operation at corner B requires $\Pel = 13.86$\,MW and at corner A, at the SCL and minimum speed, 5.88\,MW. Corner D is the lowest-power point of the envelope (5.62\,MW). At minimum speed, the power can only be reduced further by opening the discharge valve. Additionally, the header pressure sets a lower bound on the pressure ratio. Below $p_\mathrm{out}/p_1 \approx 1.48$ the compressor cannot deliver gas into the header.

\section{Flexibility Requirements}\label{sec:req}
\begin{table}[b!]
\caption{Time to full delivery of the reduction}\label{tab:req}
\centering\small
\begin{tabular}{lll}
\toprule
Service & Full delivery & Ref.\\
\midrule
Island contingency load shedding & 0.1--0.25\,s & \cite{petriccioli2021}\\
GB Dynamic Containment / Moderation & 1\,s & \cite{neso}\\
GB Dynamic Regulation & 10\,s & \cite{neso}\\
Continental Europe FCR & 30\,s & \cite{sogl}\\
Continental Europe aFRR & 5\,min & \cite{acer_afrr}\\
\bottomrule
\end{tabular}
\end{table}

\begin{figure*}[t!]
\centering
\includegraphics[width=0.65\textwidth]{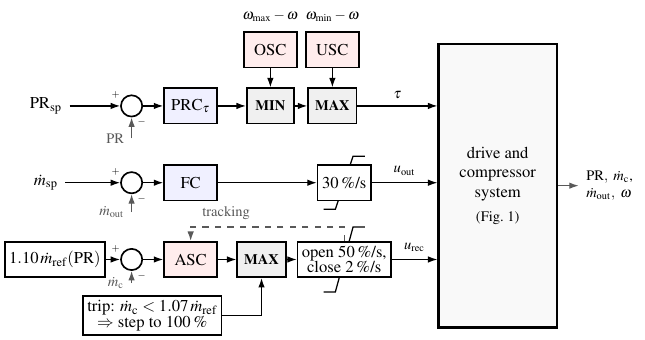}
\caption{Normal-mode control structure. Torque: pressure-ratio controller PRC$_\tau$, limited by the overspeed (OSC) and underspeed (USC) overrides. Discharge valve: flow controller FC with the actuator rate limit. Recycle valve: anti-surge controller (ASC) on the 10\,\% surge control line, with the recycle trip line at 7\,\%, fast opening and slow closing; the ASC tracks the actual valve opening. The anti-surge path is the same in power mode.}
\label{fig:normal}
\end{figure*}

In flexibility applications, how fast the power must be reduced depends on what the reduction is for. We consider two cases: grid frequency services, in which the compressor is paid to reduce its demand when the grid frequency falls, and island load shedding, in which it must reduce its demand after a generator trip. Both can be characterized by the time within which the full reduction must be delivered, summarized in Table~\ref{tab:req}.

\emph{Grid frequency services.} In Great Britain, Dynamic Containment and Dynamic Moderation require full delivery within 1\,s, and Dynamic Regulation within 10\,s~\cite{neso}. In continental Europe there are several frequency reserves, for example must frequency containment reserve (FCR) be fully activated within 30\,s~\cite{sogl}, and automatic frequency restoration reserve (aFRR) has a 5\,min full activation time~\cite{acer_afrr}. Depending on how fast a compressor can change its power consumption, it can qualify for different frequency reserve markets.

\emph{Island load shedding.} The island case is more demanding. Consider a platform supplied by gas-turbine generators. If one unit carrying $\Delta P$ trips, the remaining units of rating $S$ and inertia constant $H_\mathrm{s}$ initially see a rate of change of frequency~\cite{kundur1994}:
\begin{equation}
\frac{\mathrm d f}{\mathrm d t} = -\frac{f_0\,\Delta P}{2 H_\mathrm{s} S}.
\label{eq:rocof}
\end{equation}
For example, if a generator supplying 20\,MW trips in an islanded grid and the two remaining 25\,MW units have $H_\mathrm{s}=3$\,s, the frequency initially falls at 3.3\,Hz/s. Under-frequency protection, which disconnects loads once the frequency falls below a set threshold, would then act within a second, before the turbine governors can raise their output. Contingency-based fast load-shedding schemes therefore act on the trip signal itself and complete within 100--250\,ms~\cite{petriccioli2021}. A compressor must deliver its reduction on the same time scale, in order to stay online in such an event. The load-shedding scheme sends the total desired reduction as a single command at the moment of the trip, so the compressor sees a step change in its power reference.

\section{Control Structure}\label{sec:control}

\begin{figure*}[t]
\centering
\includegraphics[width=0.65\textwidth]{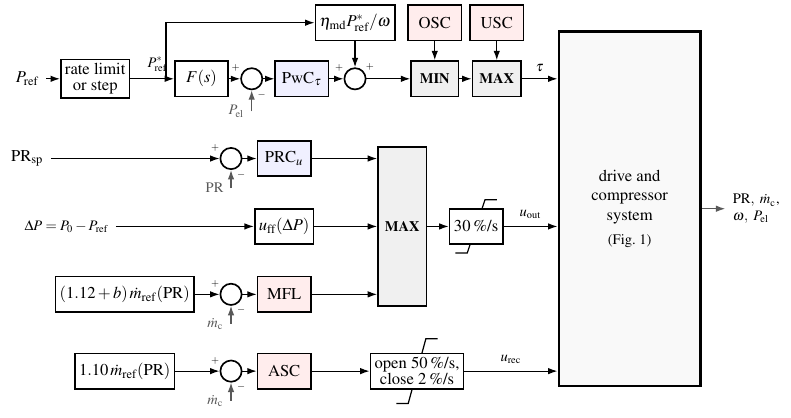}
\caption{Power-mode control structure. Torque: power controller PwC$_\tau$ around the static feedforward, limited by the overspeed (OSC) and underspeed (USC) overrides. Discharge valve: maximum of the pressure-ratio controller PRC$_u$, the valve feedforward $u_\mathrm{ff}$ and the minimum-flow limit (MFL), followed by the actuator rate limit. Recycle valve: anti-surge controller (ASC) with fast opening and slow closing; the recycle trip line (Fig.~\ref{fig:normal}) and the hold of the discharge valve during a trip (Section~\ref{sec:asc}) are not drawn.}
\label{fig:control}
\vspace{-10pt}
\end{figure*}

The structure has two modes. In normal mode the compressor follows its process set-points and in power mode it follows an electrical power reference, and the process set-points become constraints. In both modes min/max selectors combine the outputs of several controllers on one actuator. A min-selector lets the most restrictive of several controllers reduce the actuator, whereas a max-selector lets the most restrictive increase it. Every controller therefore needs anti-windup that keeps it ready to be selected.

\subsection{Normal mode}
In normal mode (Fig.~\ref{fig:normal}), two loops track the set-points: a pressure-ratio controller PRC$_\tau$ on the drive torque and a flow controller FC on the discharge valve, so that
\begin{equation}
\tau = \max\big(\mathrm{USC},\, \min(\mathrm{PRC}_\tau, \mathrm{OSC})\big), \qquad u_\mathrm{out} = \mathrm{FC},
\label{eq:normal}
\vspace{-5pt}
\end{equation}

where the overspeed and underspeed controllers OSC and USC hold the speed range. The main controllers use back-calculation anti-windup with tracking of the applied output~\cite{astrom2006}, so every selector and mode change is bumpless. The speed overrides use the external-reset form of selector override control~\cite{shinskey1996}. The integral state of OSC and USC is reset to the applied torque $\tau$ at every step, so that an inactive override outputs $\tau$ plus its proportional term. For the USC this output is $\tau - K_p(\omega - \omega_\mathrm{min})$, which lies below the applied torque by an amount proportional to the distance to the speed limit. The max-selector in \eqref{eq:normal} therefore does not select it until $\omega$ reaches $\omega_\mathrm{min}$. This matters for a fast power reduction. If back-calculation on the controller output were used instead, the output of an inactive USC would equal the applied torque. As soon as the power controller lowers the torque, the max-selector would select the USC, and the torque could then fall only as fast as the USC integrator winds down, not as fast as the power reference requests. Set-point changes in normal mode settle in 6--15\,s without opening the recycle valve.

\subsection{Anti-surge control}\label{sec:asc}
The anti-surge controller (ASC) is a PI controller that manipulates the recycle valve to keep the compressor flow to the right of the 10\,\% SCL. The valve is limited by the actuator's maximum travel rate when opening and by the 2\,\%/s limit when closing so that the operating point is not pushed back towards surge by a fast closure. As common in commercial anti-surge systems, a recycle trip line at a 7\,\% flow margin to the reference surge line is also implemented behind the PI loop (Fig.~\ref{fig:normal}). When the measured flow crosses this line, the valve command steps to fully open. The ASC output tracks the actual valve opening, so after a trip the valve closes at the slow-closing rate until the anti-surge PI controller output again becomes the active constraint at the SCL. Without this tracking the PI output stays at zero during the slow closure, the valve closes through the SCL band, and the 7\,\% trip line is triggered again. When a trip is activated, the current discharge valve opening becomes a constraint. While a trip is active, the discharge valve is held at or above this constraint, so that the process controls do not close it while the recycle valve is open. To check that the trip line does not fire on ordinary disturbances, the suction valve is closed to 35\,\% in 2\,s in normal mode. The surge margin falls to 7.9\,\%, below the SCL but above the trip line; the PI controller ensures the recycle valve goes to to 2.8\,\% and the trip line is not reached.

\subsection{Power mode}
When a power reference $\Pref$ is issued, the flow set-point is dropped and the structure of Fig.~\ref{fig:control} is used. The reference is passed as $P^\ast_\mathrm{ref}$, either as a step or through a rate limiter . The torque is controlled with the power controller PwC$_\tau$ instead of the pressure ratio controller, PRC$_\tau$. $\Pel$ is a static function of torque, so the torque is computed by inverting Equation \eqref{eq:pel},
\begin{equation}
\tau = \max\big(\mathrm{USC},\, \min(\eta_\mathrm{md}P^\ast_\mathrm{ref}/\omega + \mathrm{PwC}_\tau,\, \mathrm{OSC})\big),
\label{eq:torque}
\end{equation}
This way, the feedforward sets almost all of the torque and PwC$_\tau$ only corrects model error. The measured power lags the actual power because of the measurement filter (and, where present, the drive lag). PwC$_\tau$ therefore compares $\Pel$ not with $P^\ast_\mathrm{ref}$ but with $F(s)P^\ast_\mathrm{ref}$, where $F(s)$ is the same lag with one sample of delay, so that both signals are delayed equally. Without $F$, PwC$_\tau$ would treat the lag as an error during a fast power reduction, and the power would undershoot by up to 1\,MW. In normal mode the feedforward uses the measured power, so PwC$_\tau$ remains at zero and the switch to power mode is bumpless.

Because flow is no longer a control objective, the discharge valve is used to keep the pressure ratio as high as the power allows:
\vspace{-2mm}
\begin{equation}
u_\mathrm{out} = R_{30}\big(\max(\mathrm{PRC}_u,\, u_\mathrm{ff},\, \mathrm{MFL})\big),
\label{eq:valve}
\end{equation}
where $R_{30}$ is the 30\,\%/s actuator rate limit and all controllers track the actual valve position. The reverse-acting PRC$_u$ closes the discharge valve to hold the pressure-ratio set-point. Whereas, the minimum-flow limit controller (MFL) opens it to keep $\mc \ge (1.12 + b)\,\dot m_\mathrm{ref}(\mathrm{PR})$, i.e.\ 2\,\% to the right of the SCL plus a dynamic margin $b$. With falling power the operating point moves down and to the left in the map, Fig.~\ref{fig:envelope}. Once it reaches the MFL line, the MFL controller opens the discharge valve, which lowers the pressure ratio and lets the flow stay at the line while the power continues to fall.

%------ This paragraph could be or not be shortened IMO -----
The gap between the MFL line and the SCL is deliberate. Without it, the MFL and the ASC would hold the same compressor flow $\mc = \dot m_\mathrm{out} + \dot m_\mathrm{rec}$ at the same value with different valves, so any split between delivery and recycle would satisfy both. Since the ASC opens fast and closes slowly, the recycle valve would then open a little further after every disturbance, and the drive would spend its power recirculating gas.
%Flow is transferred from delivery to recycle while the power stays at its reference: the drive spends its power recirculating gas. The same transfer is triggered when a fast transient pushes the operating point into the ASC band, and in its extreme form when the trip line opens the recycle valve fully (Section~\ref{sec:results}).
%---------------------------------------------------------

Because the MFL acts only once the operating point reaches its line, it lags a fast power reduction. So, the dynamic margin $b$ and the valve feedforward $u_\mathrm{ff}$ are anticipatory.
\begin{align}
b &= \min\big(0.2,\, \max(0.02,\, T_b\, r)\big), \qquad T_b = 0.6\,\mathrm{s}, \label{eq:bias}\\
u_\mathrm{ff} &= 0.8\, \mathrm{sat}_{[0,1]}\!\left(\frac{\Delta P - 0.1}{0.5 - 0.1}\right). \label{eq:uff}
\end{align}
In \eqref{eq:bias}, $r$ is the ramp rate of $P^\ast_\mathrm{ref}$ in per unit of the 15\,MW rating per second, and $b = 0.2$ for a step. The margin shifts the MFL line to the right in proportion to the ramp rate, so the valve starts opening earlier for a faster reduction. In \eqref{eq:uff}, $\Delta P = (P_0 - \Pref)/P_\mathrm{rated}$ is the size of the requested reduction, with $P_0$ the power when the request arrives, so the feedforward opens the valve in proportion to how much power is to be shed. Through the max-selector, $u_\mathrm{ff}$ can only open the valve further. For small reductions it stays below the current opening and the MFL does the work. For large ones it does most of the opening before the operating point reaches the MFL line. Both terms are held while the reference falls and then released with a 5\,s time constant, after which PRC$_u$ and the MFL settle the valve. The MFL gain must be carefully tuned as it participates in a lightly damped mode of 0.7\,Hz.

\subsection{Minimum power, supervisor and return to normal mode}
In power mode only the torque acts on the power; the discharge valve only keeps the flow at or to the right of the MFL. The lowest power reachable with \eqref{eq:torque}--\eqref{eq:valve} is therefore where the MFL line meets minimum speed, $P_\mathrm{min} = 5.87$\,MW, around corner A. There the USC holds the torque, and no controller opens the discharge valve to lower the power further towards corner D (Section~\ref{sec:discussion}).The supervisor therefore clips references between $0.85\,P_\mathrm{min}$ and $P_\mathrm{min}$ to $P_\mathrm{min}$, delivers $P_\mathrm{min}$ and reports the shortfall, and refuses lower references by shutting the machine down (torque zero, recycle open, discharge valve closed, all at their rate limits). An online monitor does the same if the reference cannot be met with no lever left. When the request ends, the machine returns to normal mode with the torque increase limited to 0.1\,p.u./s and the discharge-valve closing to 5\,\%/s, which avoids chatter between PRC$_\tau$ and the OSC while the pressure ratio recovers.
\vspace{-1mm}
\section{Results}\label{sec:results}

\subsection{Test points and nominal results}
Five start points S1--S5 were taken on the $\Pel = 12$\,MW iso-power line, from the SCL (S1, PR 1.935, margin 11.2\,\%) to the fully open valve (S5, PR 1.58). Along this line the speed varies between 11\,400 and 11\,720\,rpm, and the start points differ mainly in how the 12\,MW is split between pressure ratio and flow. From S4 and S5 every power reduction considered here, including a step to 6\,MW, stays at least 12.5\,\% from surge. The results below therefore use only the three points closest to surge, S1 (PR 1.935, 11.2\,\%), S2 (PR 1.90, 20.9\,\%) and S3 (PR 1.80, 44.4\,\%), and a 6\,MW reduction target, a 50\,\% power reduction that ends 40\,rpm above minimum speed. Each run starts in normal mode at its own set-points.

\begin{figure*}[t]
\centering
\includegraphics[width=0.95\textwidth]{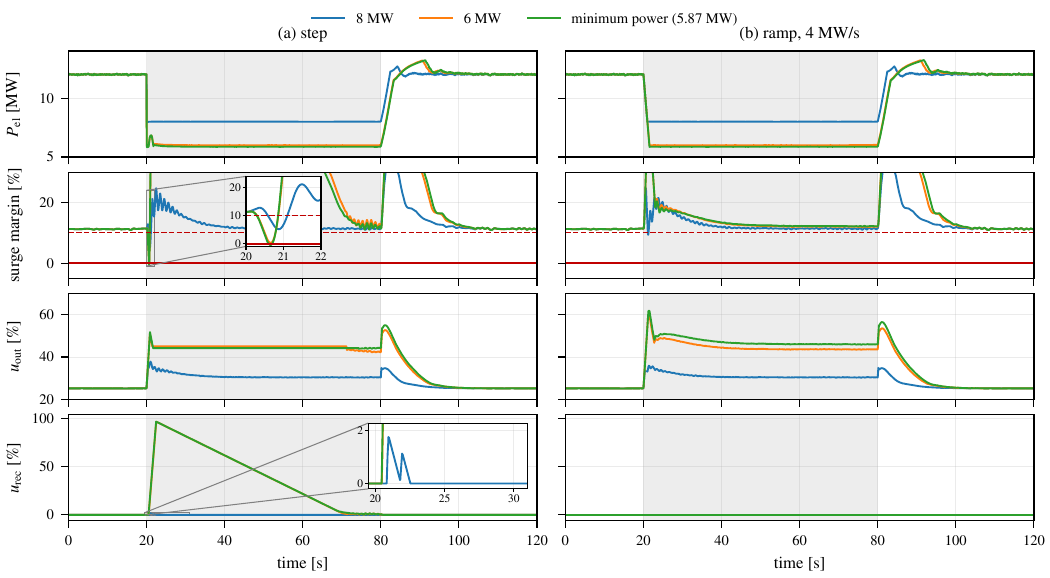}
\caption{Requests of 8\,MW, 6\,MW and minimum power (5.87\,MW) from S1 between 20 and 80\,s (shaded), (a) as a step and (b) with a 4\,MW/s rate limit, followed by the return to normal mode. Dashed red: SCL (10\,\%); solid red: surge line. The surge margin is truncated at 30\,\%: the 6\,MW and minimum-power steps, which trip the recycle valve open, leave the plot until the valve closes. Inset: recycle opening for the 8\,MW step.}
\label{fig:nominal}
\end{figure*}

Fig.~\ref{fig:nominal} shows requests of 8\,MW, 6\,MW and $P_\mathrm{min}$ from S1, delivered as a step (a) and with a 4\,MW/s rate limit (b), and the return to normal mode after 60\,s. The rows show the electrical power, the surge margin, and the openings of the discharge and recycle valves.

With the rate limit, all three requests are delivered without recycle flow and with a minimum margin of 9.3--11.3\,\%, which is acceptable operation. The machine settles on the MFL line: 8\,MW at PR 1.729 and 10\,480\,rpm, 6\,MW at PR 1.580 and 9\,940\,rpm with the discharge valve at 44\,\%, and $P_\mathrm{min}$ at PR 1.570 and minimum speed with the valve at 46\,\%. The valve feedforward opens the discharge valve to 60\,\% at the start of large power reductions, so the margin actually rises after the initial drop during the power reduction, however at the price of a pressure ratio that falls ahead of the power.

The step reductions deliver the same power reductions as in the rate-limited case quicker, but the anti-surge valve opens in every case to varying extents. For the reduction to 8\,MW the PI loop only opens the recycle valve by 1.8\,\% for 2\,s (inset of Fig.~\ref{fig:nominal}a), and a lightly damped 0.7\,Hz oscillation decays within 10\,s. For 6\,MW and $P_\mathrm{min}$ the operating point reaches the surge line, the recycle trip line fires within the first second and the recycle valve opens almost fully. The drive still meets the power target, but the power is spent recirculating gas as the pressure ratio falls below the 1.48 needed to deliver into the header. Gas delivery stops for 28\,s although the discharge valve is held open. As the recycle valve closes at 2\,\%/s, delivery resumes. Since the ASC tracks the closing valve, the trip line does not fire again, and after 50\,s the machine is back at the same operating point as with the rate limit. This interruption of delivery is the price of a step from a point at the SCL. In both columns the return to normal mode takes 20\,s and is limited by the torque rate limit.

\subsection{Rate of the power reduction}
\begin{figure}[b!]
\centering
\includegraphics[width=\columnwidth]{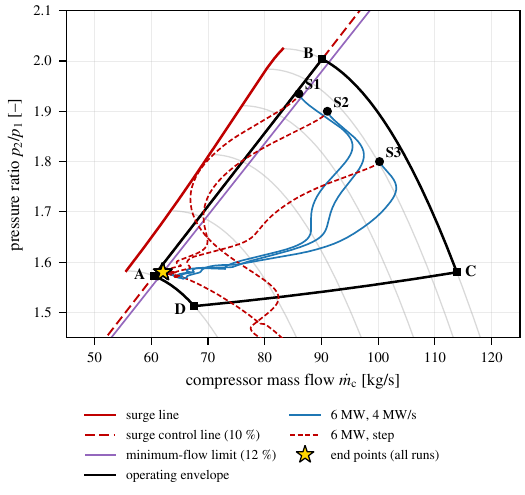}
\caption{Paths in the compressor map of 6\,MW requests from S1--S3, with a 4\,MW/s rate limit and as a step. The step from S1 triggers the recycle trip and leaves the plotted region below PR 1.45 before returning. Star: end point of all runs, 6\,MW on the minimum-flow limit (PR 1.58). Grey: speed lines.}
\label{fig:paths}
\end{figure}

Table~\ref{tab:rate} gives, for the 6\,MW request from S1--S3, the time to 90\,\% of the reduction and the minimum surge margin and peak recycle opening for rate limits from 1 to 10\,MW/s and for a step. Fig.~\ref{fig:paths} shows the paths of the 4\,MW/s and step runs in the compressor map.

Every rate limit up to 10\,MW/s is delivered from all three start points without recycle, with a minimum margin of 11.3--11.9\,\% and a pressure ratio that never falls below 1.567 (final 1.580). At 10\,MW/s, 90\,\% of the 6\,MW reduction is delivered in 0.54\,s. The minimum margin is the same at 10\,MW/s as at 1\,MW/s since during the power reduction the feedforward and the dynamic margin keep the operating point well to the right of the MFL line. So, the minimum is reached only at the end, when the machine settles on the MFL line at 6\,MW. The step gains half a second over 10\,MW/s. From S3 it is also clean but from S2 it dips 2\,\% into the SCL band for 0.3\,s without opening the recycle, and from S1 it comes within 0.3\,\% of the surge line and triggers the recycle trip described above.

The mechanism is visible in Fig.~\ref{fig:paths}. On a step, the electrical power and the drive torque fall instantly, and the rotor decelerates at a rate set by the power difference whereas the inertia and the head falls with the speed. The discharge valve, however, needs 1.2\,s to stroke from its initial 25\,\% to the 60\,\% asked for by the feedforward, so during the first second the discharge pressure cannot be relieved as fast as the head falls, and the operating point moves left, towards surge. With a rate limit the deceleration is spread over the power reduction, the valve keeps ahead of it, and the operating point first moves to the right of its start and then down to the MFL line. The fastest safe power cut from a point near surge is therefore set by the stroking rate of the discharge valve, not by the drive. The higher the start pressure ratio, the more pressure has to be relieved and the closer the step comes to surge, which is why the step is clean from S3, marginal from S2 and trips from S1.

\begin{table}[t]
\caption{6\,MW request from S1--S3: rate limit, dynamic margin, time to 90\,\% of the reduction, minimum surge margin and peak recycle opening}\label{tab:rate}
\centering\small
\setlength{\tabcolsep}{3.2pt}
\begin{tabular}{lcccccc}
\toprule
 & & & \multicolumn{3}{c}{$\min\mu$ [\%]} & $\max u_\mathrm{rec}$\\
\cmidrule(lr){4-6}
Rate & $b$ & $t_{90}$ [s] & S1 & S2 & S3 & [\%]\\
\midrule
1\,MW/s  & 4\,\%  & 5.40 & 11.3 & 11.9 & 11.9 & 0\\
2\,MW/s  & 8\,\%  & 2.70 & 11.3 & 11.9 & 11.9 & 0\\
4\,MW/s  & 16\,\% & 1.35 & 11.3 & 11.9 & 11.9 & 0\\
6\,MW/s  & 20\,\% & 0.90 & 11.3 & 11.9 & 11.9 & 0\\
10\,MW/s & 20\,\% & 0.54 & 11.3 & 11.9 & 11.9 & 0\\
step     & 20\,\% & 0.01 & 0.3  & 7.0  & 11.9 & 97 (S1, trip)\\
\bottomrule
\multicolumn{7}{l}{\footnotesize Rates in MW/s of the 15\,MW rating; 1\,MW/s $=6.7$\,\%/s. $t_{90}$ from S1.}
\vspace{-1mm}
\end{tabular}
\end{table}

% \begin{figure*}[t]
% \centering
% \includegraphics[width=\textwidth]{fig_rate.pdf}
% \caption{6\,MW request from (a) S1, (b) S2 and (c) S3 at 20\,s, with a 4\,MW/s rate limit and as a step. Dashed: SCL (10\,\%); solid: surge line.}
% \label{fig:rate}
% \end{figure*}

In the step from S1 and S2, the power also rebounds, by 0.8 and 0.5\,MW, within the first second and returns to the target within 2.4\,s (Fig.~\ref{fig:nominal}a). The rebound is caused by the USC. As the minimum speed is approached, the gap between the USC output and the applied torque closes, and the max-selector hands the torque to the USC. The USC then raises the torque to stop the deceleration at minimum speed, and the power rises above its target until the speed settles. It shapes the approach to minimum speed and limits the undershoot of the speed, but on an island it means that part of the shed load returns briefly. With rate limits of 6 and 10\,MW/s a smaller rebound of 0.2--0.4\,MW remains from S1.

\subsection{Drive lag and signalling delay sensitivity}\label{sec:lag}
For the island case the drive is modelled with the lag $\tau_\mathrm{e}$ between torque command and air-gap torque, and the request with a dead time $\theta$ for detection and communication. Table~\ref{tab:island} gives the time to 90\,\% of the 6\,MW reduction from S1, with the minimum margin and peak recycle. With a step, even $\tau_\mathrm{e}=100$\,ms and $\theta=100$\,ms deliver 90\,\% of the reduction in 0.33\,s, close to the 100--250\,ms window of Table~\ref{tab:req}, and 4.7\,MW of the 6\,MW within 0.25\,s. The drive lag acts as a filter on the step so the torque falls over $3\tau_\mathrm{e}$ rather than instantly, which reduces the deceleration at the start of the power reduction and gives the discharge valve time to open. Without drive lag the step triggers the recycle trip as in Fig.~\ref{fig:nominal}a. With $\tau_\mathrm{e}=50$\,ms the operating point stays clear of the trip line and the recycle opens by 3\,\% under PI action. With $\tau_\mathrm{e}=100$\,ms the recycle stays closed (minimum margin 4.4\,\%). The signalling delay only shifts the response in time. With the 4\,MW/s rate limit, reducing the power takes 1.35--1.55\,s and keeps a margin of 11.2\,\% for all combinations and with a 10\,MW/s limit reaches 90\,\% in 0.54\,s.

\begin{table}[b!]
\caption{6\,MW request from S1: time to 90\,\% of the reduction [s] / minimum surge margin [\%] / peak recycle [\%]}\label{tab:island}
\centering\footnotesize
\setlength{\tabcolsep}{3.0pt}
\begin{tabular}{llccc}
\toprule
 & & $\tau_\mathrm{e}=0$ & 50\,ms & 100\,ms\\
\midrule
\multirow{2}{*}{step} & $\theta = 0$ & 0.01/0.3/97 & 0.12/1.6/3.2 & 0.23/4.5/0\\
 & $\theta = 100$\,ms & 0.11/0.3/97 & 0.22/1.6/3.3 & 0.33/4.4/0\\
\midrule
\multirow{2}{*}{4\,MW/s} & $\theta = 0$ & 1.35/11.3/0 & 1.40/11.3/0 & 1.45/11.3/0\\
 & $\theta = 100$\,ms & 1.45/11.2/0 & 1.50/11.3/0 & 1.55/11.3/0\\
\bottomrule
\end{tabular}
\end{table}

\subsection{Rotor inertia sensitivity}
By the torque balance in \eqref{eq:rotor}, the inertia sets the deceleration that follows a given torque reduction. For the same power step, a rotor with half the inertia loses speed, and head, twice as fast. Table~\ref{tab:inertia} repeats the 6\,MW request from S1 for $H =$ 0.5, 1 and 2\,s with the same tuning. The 4\,MW/s rate limit is insensitive to the inertia as the ramp, not the rotor, sets the deceleration, and the valve keeps ahead of it in all three cases. With a step, a heavier rotor ($H=2$\,s) keeps the recycle closed whereas a lighter one ($H=0.5$\,s) crosses the surge line by 20\,\% and triggers the recycle trip. The recycle valve then stays fully open, and delivery does not resume within the 60\,s request. The same happens at 10\,MW/s. The rate limit therefore has to be chosen with the inertia. So, the dynamic margin \eqref{eq:bias} and the valve stroking rate were calibrated for $H=1$\,s.

\begin{table}[b!]
\caption{6\,MW request from S1, inertia constant $H$: minimum surge margin [\%] / peak recycle [\%]}\label{tab:inertia}
\centering\small
\begin{tabular}{lccc}
\toprule
 & 4\,MW/s & 10\,MW/s & step\\
\midrule
$H=0.5$\,s & 11.0/0 & $-4.2$/100 & $-20.3$/100\\
$H=1$\,s   & 11.3/0 & 11.3/0 & 0.3/97\\
$H=2$\,s   & 11.3/0 & 11.3/0 & 11.3/0\\
\bottomrule
\end{tabular}
\end{table}

\section{Discussion}\label{sec:discussion}

The results separate two settings. For grid services a rate limit with the scaled margin is sufficient. Dynamic Containment asks for full delivery within 1\,s and no faster than 0.5\,s, which a 6--10\,MW/s limit meets for a 6\,MW reduction of power consumption without recycle from any start point. For island load shedding the request is a step and the window is 100--250\,ms. The step meets it, but from start points at the SCL it triggers the recycle trip, unless the drive itself provides a lag of 50--100\,ms. A 10\,MW/s rate limit, which reaches 90\,\% of the power reduction in half a second, avoids the trip from every start point. Whether this is fast enough depends on how much of the shed load the island needs within the first few hundred milliseconds.

In the step from S1 the compressor stays online, holds the requested power throughout, and ends at the same operating point as with the rate limit (Fig.~\ref{fig:paths}). From the point of view of the machine and of the power system, the request has been met. From the point of view of the process it has not as delivery into the header stops for 28\,s, and full delivery is restored only after 50\,s. Whether this is acceptable depends on process requirements. A process that relies on continuous delivery, such as a gas-injection or export system with tight pressure limits, may treat a 28\,s interruption as equivalent to a shutdown of the compressor, even though the machine is running and recovers automatically. In such a case the step response from points near the SCL should be classified as a production loss, and the rate limit, or a start point further from the SCL, is the safer choice.

With the torque feedforward the electrical response is essentially immediate. What remains is to relieve the discharge pressure as the head falls with speed. A faster-stroking discharge valve would raise the safe rate. So would opening the anti-surge valve deliberately as a fast vent, but this is effectively what the trip line does in the step from S1, and it interrupts delivery.

Below $P_\mathrm{min}$, the power can be reduced further by holding the speed at its minimum and opening the discharge valve along the minimum-speed line towards corner D. A second stage that does this with a reverse-acting power controller on the discharge valve is implemented in the accompanying code but not shown here. It lowers the reachable power from 5.87 to 5.62\,MW, a further 1.7\,\% of rating, but at a pressure ratio of 1.51 instead of 1.57, and in tens of seconds rather than the fraction of a second of the speed path. For fast flexibility the gain is small though it may be worthwhile for slow services.

\emph{Limitations.} The map left of the surge peak and the stand-still behaviour are estimated, so behaviour after a surge-line crossing, such as the $H=0.5$\,s cases, is less certain than the crossing itself; the plena are isothermal; the drive and interface to the power grid are represented by a lag; the trip-line logic is a simplified version of commercial implementations; and the tuning was obtained for one machine. The 0.7\,Hz mode that limits the MFL gain, which we attribute to the interaction of valve opening, compressor flow and rotor speed under constant electrical power, deserves a linear analysis, and the rebound after a step calls for an approach to minimum speed that does not return load to the grid.

\section{Conclusions}

We demonstrate that an electrically driven centrifugal compressor can cut its power by 40\,\% of its rating within half a second without opening its anti-surge valve, provided the cut is rate-limited and the discharge valve is moved ahead of the operating point. A static power feedforward makes the electrical response essentially immediate. A feedforward of the discharge valve scaled with the size of the cut, and a minimum-flow limit with a dynamic margin scaled with the ramp rate, keep the operating point away from surge. From the start points closest to surge, rate limits of up to 10\,MW/s are delivered with an unchanged minimum margin, whereas a step from the surge control line triggers the recycle trip. The machine holds the power and recovers on its own, but gas delivery is interrupted for 28\,s, which some processes will treat as a shutdown. Further work will analyse the 0.7\,Hz mode of the valve loops, shape the approach to minimum speed to remove the rebound, compare the regulatory structure with economic MPC, and validate the scheme on a test rig.

The model, the controllers and the scripts that reproduce all figures and tables are available at {\footnotesize\url{https://github.com/AISL-at-Imperial-College-London/flex-comp}}.
% ---------------------------------------------------------------- references
% NOTE: bibliographic details to be verified before submission.

\end{document}